\documentclass[aps,jmp,groupedaddress,twocolumn,showkeys]{revtex4}
\usepackage[utf8]{inputenc}
\usepackage{amssymb}
\usepackage[T1]{fontenc}
\usepackage{hyperref}

\begin{document}

\title{Comment on ‘Quantum principle of relativity’}

\author{Ryszard Horodecki}

\address{International Centre for Theory of Quantum Technologies, University of Gdańsk, Jana Bażyńskiego 1A, 80-308 Gdańsk, Poland}
\address{National Quantum Information Center of Gdańsk, Wita Stwosza 53, 80-308 Gdańsk, Poland}


\date{\today}

\begin{abstract} \small
  Dragan and Ekert in the paper (2020 \emph{New. J. Phys.} \textbf{22} 033038) presented
  `quantum principle of relativity' (QPR) based on Galilean
  principle of relativity, which involves both superluminal $G_S$ and
  subluminal $G_s$ families of observers and argue that then they are
  considered on the same footing it `implies the emergence of
  non-deterministic dynamics, together with complex probability
  amplitudes and multiple trajectories.'. Here we discuss QPR in the
  context of Heisenberg’s classification of the fundamental physical
  theoretical models under the role universal constants of nature:
  Planck’s constant $h$ and speed of light $c$. We point out
  that both the superluminal
  and subluminal branches are separable in the sense that there is no
  mathematical coherent formalism that connect both branches.
  This, in particular, implies that the QPR is incomplete.
\end{abstract}

\maketitle

\section{Introduction}

In their inspiring paper \cite{QPR}, the authors proposed a `quantum principle of relativity' (QPR) based on Galilean principle of relativity, which involves branches of coordinate transformations corresponding to subluminal $G_s$ and superluminal (tachyonic) $G_S$ families of observers. The authors claim that  taking into account both branches $G_s$ and $G_S$ on the same footing implies non-deterministic decays, necessity for quantum superpositions and complex probability amplitudes.

Idea of tachyons as new particles associated with the superluminal $G_S$ branch \cite{BILDS} has a long history and was developed in different directions. It involves the extension of Minkowski space-time to the pseudo-Euclidean space-time \cite{March_Antippa,Parker,Olekhowsky_Rec,Antippa_Ev,Recami_Mig,Antippa_Nuovo,Antippa_Phys,Maccarrone_Recami}, a quantum field theory of tachyons \cite{Feinberg}, the model of pseudotachyons \cite{H_wave-particle,H_PLA,H_Extended_wave,Molski}, causal model of tachyons \cite{Rembielinski}. 

One of the reasons for the interest in tachyons was the overwhelming success Salam–Weinberg  model (see for instance \cite{H_PLA, Higgs1, Higgs2, Dragan_et_all}) where the gauge symmetry is spontaneously broken by filling the vacuum with Higgs field particles which can be considered formally as  tachyons which have been converted into subluminal particles. It is very intriguing,  that even though Einstein’s ban on faster-than-light speeds does not prohibit the existence of superluminal particles (tachyons) so far still there are no convincing experiments for the existence of such exotic particles \cite{Ehrlich}.

In contrast to all the above approaches, the paper \cite{QPR} seems to open a new horizon, indicating a possible relationship between QPR and quantum mechanics. The latter though works perfectly and is reliable in an appropriate regime, however leaves us with its central enigma: The phenomenon of quantum randomness `something that happens without  any cause goes against our rational understanding of reality'\cite{QPR}. In \cite{QPR} the authors   argue, in particular,  that non-deterministic quantum dynamics is a consequence of QPR.

However, the analysis of possible implications of QPR shows that there are some gaps in the authors' reasoning, which may raise doubts about the correct interpretation of the results.

\section{Operational separability of subluminal and superluminal Galilean branches}

In section 4, the authors derive the concept of probability-like quantities by postulating a relativistic invariant as a proper time characterizing freely moving point-like  particle along single path in the $G_s$  branch (eq. (11)). To make the relativistic invariant dimensionless they choose the  constant $\frac{2\pi}{h}$ as the constant of proportionality. This seemingly innocent step is crucial as the constant $h=2\pi\hslash$ ($\hslash$ Dirac constant) introduced by Planck  played the role of a concept synthesizer in the development of quantum theory from Einstein---de Broglie wave-particle relations to the concept of quantum information \cite{Brukner,Stuckey,Horodecki_QI}

As Heisenberg has already noticed \cite{Heisenberg}, the introduction of the universal constants $\hslash$ and $c$ is closely related to the step changes in the paradigms of basic theoretical models in physics: (i) the equations $\frac{1}{c} = \hslash =0$ characterize the non-relativistic, classical model, (ii) $\frac1{c} =0$, $\hslash\neq  0$: the non-relativistic quantum model, (iii) $\frac1{c} \neq  0$, $\hslash=0$ the relativistic classical model, (iv) $\frac1{c} \neq 0$ and $\hslash\neq 0$ the relativistic, quantum model. 

In the above context, a great peculiarity is the distinguished role of the Galileo's space-time $G_4$ in the non-relativistic quantum model $\frac1{c}=0$, $\hslash\neq 0$, were constant $\hslash$ is responsible for Heisenberg non-commutative algebra of measurement operators, which imposes universal constraints understood as a system of the possibility of extracting information from a quantum system.

Remarkably the non-relativistic quantum model (ii) introduces inherent randomness consistent with the $G_s$  and reproduces the extent of stable macroscopic measures with a hierarchical structure (quarks, nucleons, nuclei, atoms, molecules\ldots). 

However QPR employ $G_2$ ((1+1)-dimensional) space-time. Here the authors assume from outset that the relativistic invariant they are looking for must behave like probability, then they write down conditions (12-14) and then notice that there is a special case of such a function (15). However, this by no means shows that the inclusion of superluminal observers `implies the emergence of non-deterministic dynamics, together with complex probability amplitudes' \cite{QPR}.

In fact, the authors introduce a link between special relativity and quantum theory only on the level of subluminal branch $G_s$, restricting  their consideration to a freely moving point-like particle propagating along multiple path.
In other words, apart from intuition, there is no coherent formalism that treats both branches of $G_S$ and $G_s$ together.

Thus, even on a conceptual level, it will be a great challenge basing on QPR to justify the double  role of tachyons, which would explain non-deterministic decays, quantum randomness and at the same time the unquestionable stability of certain quantum structures. 

In this sense keeping both branches at once has no operational significance. Indeed, for QPR the existence of the universal constant $c$ is crucial, in the contrast to the $\hslash$, which applies only to the branch $G_s$ and it does not \emph{a priori} any connection with $c$. In particular it seems unlikely that this asymmetry can be removed by any extension of QPR to include the relationship between $\hslash$ and $c$. (The combination $\frac{\hslash}{c} \neq 0$ appears already in Einstein's wave-particle relations as well in the Bethe--Salpeter equation \cite{Bethe}, which however encounters a number of mathematical and interpretative difficulties). 

To overcome the formal separation between the branches $G_S$ and $G_s$ the authors develop an attractive interpretative footbridge which involves analogy between the indeterministic behavior of superluminal (subluminal) particles, non-classical motion of particles and non-classical properties of quantum particles including the wave properties in the context of Huygens principle \cite{QPR} (see e.g. \cite{Path_integral}).

\section{Tachyonic sources of quantum randomness?}

It is well known that the quantum randomness has  two faces: measurement-like  and quantum decay process.

The first  type of quantum randomness  manifests itself in measurements, while quantum states evolve deterministically.  This type of randomness is contextual and so far we do not have any consistent theory of measurement. We do not know whether it is the result of spontaneous breaking of unitary evolution due to the existence nonunitary reduction process $R$ or the result of the irreversible process in the thermodynamic limit. All we rely on is the von Neumann reduction postulate and Born's rule. It is not entirely clear whether the later obliges in any scenarios allowed by QPR. The question arises here: How will the measurement phenomenon (in a given measurement context) be seen by the observer from the superluminal system and how from the subluminal one.

The second kind of quantum randomness is due to decaying of particles and the nature of metric  time, where the evolution of states is usually described in terms dynamical semigroup \cite{Hiesmayr}. So far we have no convincing proof that the measurement randomness can always be reduced to the randomness associated with particle decays. This would then require a separate analysis. 
However it seems unlikely that both of the above types of quantum randomness have their common origin in tachyon indeterminism.

\section{Final remarks}

In any theory, measurable quantities should be identified. This is especially true of the 1+3 dimensional space-time generalization, which introduces a three-dimensional time vector. In this case the subluminal and superluminal observers cannot be treated on the same footing and question arises whether the modulus of the three-time vector is measurable.

If we were to take QPR seriously as a new physical postulate then all laws including Born's rule should be valid in any frame of reference within reasonable limits, but it  unlikely us as the superluminal extension of special relativity in 1+3 dimensional space-time is not covariant \cite{Dragan_et_all}. 

According to the above arguments, the QPR cannot be treated as complete in the sense that it is neither universal nor formal. Einstein was convinced that `only the discovery of a universal formal principle could lead us to assured results.' \cite{Einstein}. In particular it does not provide any measurable and potentially observable effects. Nevertheless, the idea to consider both branches of $G_S$ and $G_s$ on the same level opens up a new way of thinking about possible relations between special relativity and quantum mechanics.

Note that there is a fundamental asymmetry between QPR and probability amplitudes concept due to the status of constants  $c$ and $\hslash$. Namely for QPR the existence of a universal (`critical') constant $c$ is crucial for both superluminal and subluminal branches, while Planck's constant enters only as a proportionality factor $\frac{2\pi}{h}$ in the subluminal branch. To restore symmetry, in the latter, this seems reasonable to adopt the following postulate: 

\emph{The value of Planck's constant $h$ measured in any inertial system $G_s$ always takes the same value}.

It is compatible with Galileo-Einstein principle: `no preferred reference frame', which includes the SO(3) (subgroup of Lorentz and Galileo's transformations) invariance of measurements of h between different reference frames of mutually complementary spin measurements \cite{Stuckey}. The above postulate has some experimental support as that the Stern-Gerlach experiment  realizes the measurement of `a universal constant of  nature Planck’s constant' \cite{Weinberg,Stuckey}.

Of course, such symmetrization within the $G_s$ branch does not make any footbridge between the two  families of inertial  observers. Nevertheless, the restored on this level symmetry underlines basic origin of the relativistic constraint on the possible form of probability amplitudes. However, the physical consequences of the restored symmetry will require in-depth research.

In conclusion, the analysis of QPR  in the context of Heisenberg’s classification of the fundamental physical models shows  the unequal footing  of $G_S$ and $G_s$ branches which are operationally separable. Thus the inclusion of superluminal observers does not imply the emergence of quantum features at the level of the subluminal branch $G_s$. Hence interpretative statements \cite{QPR} that the fundamental features of quantum theory can be derived from Galilean principle of relativity do not have a coherent conceptual and formal basis. 

Finally, it should be noted that the original paper \cite{QPR} has been criticized earlier along various lines \cite{Grudka, Santo}. Grudka and Wójcik \cite{Grudka} showed that there is no need to introduce superluminal observers, because the antisymmetric branch of the transformation considered in \cite{QPR} can appear as a consequence of changing the coordinate names. Del Santo and Horvat  \cite{Santo} more broadly criticize Dragan and Ekert's positions and provide a number of counter-arguments against the claim  that the quantum indeterminism and the principle of superposition can be derived solely derived from the Galilean principle of relativity. 

\section*{Acknowledgement}

I thank A Dragan for earlier correspondence,
M Eckstein and T Miller for critical reading of the manuscript and valuable comments, and P Horodecki and K Horodecki for helpful comments.
The work was supported by the Foundation for Polish Science (IRAP project, ICTQT, contract No. MAB/2018/5, co-financed by EU within Smart Growth Operational Programme).


\begin{thebibliography}{10}

  \bibitem{QPR} Dragan A and Ekert A 2022 \emph{New. J. Phys.} \textbf{22} 033038
  \bibitem{BILDS} Bilaniuk O M P, Deshpande V K and Sudarshan E C G  1962 \emph{Am. J. Phys.} \textbf{30} 718
  \bibitem{March_Antippa} Marchildon L, Antippa A F and Everett A 1983 \emph{Can. J. Phys.} \textbf{61} 256
  \bibitem{Parker} Parker L 1969 \emph{Phys. Rev.} \textbf{188} 2287
  \bibitem{Olekhowsky_Rec} Olekhowsky V S and Recami E 1971 \emph{Lett. Nuovo Cimento} \textbf{1} 165
  \bibitem{Antippa_Ev} Antippa A F and Everett A E  1971 \emph{Phys. Rev. D} \textbf{4} 2198 Antippa A F Everett A E 1973 \emph{Phys. Rev. D} \textbf{8} 2352
  \bibitem{Recami_Mig} Recami E and Mignani R 1972 \emph{Lett. Nuovo Cimento} \textbf{4} 144 Recami E and Mignani R 1973 \emph{Lett. Nuovo Cimento} \textbf{8} 110
  \bibitem{Antippa_Nuovo} Antippa A F 1972 \emph{Nuovo Cimento A} \textbf{10} 389
  \bibitem{Antippa_Phys} Antippa A F 1975 \emph{Phys. Rev D} \textbf{11} 724
  \bibitem{Maccarrone_Recami} Maccarone G D and Recami E 1982 \emph{Lett. Nuovo Cimento} \textbf{34} 231
  \bibitem{Feinberg} Feinberg G 1967 \emph{Phys. Rev.} \textbf{159} 1089
  \bibitem{H_wave-particle} Horodecki R 1984  \emph{Il Novo Cimento B} \textbf{80} 217
  \bibitem{H_PLA} Horodecki R 1988 \emph{Phys. Lett. A} \textbf{133} 179
  \bibitem{H_Extended_wave} Horodecki R 1988 \emph{Il Nuovo Cimento} \textbf{102} 27
  \bibitem{Molski} Molski M 2006 \emph{Eur. Phys. J.} \textbf{40} 411
  \bibitem{Rembielinski} Rembieliński J 1997 \emph{Int. J. Mod. Phys. A} \textbf{12} 1677
  \bibitem{Higgs1} Ramazanoğlu F M 2018 \emph{Phys. Rev. D} \textbf{98} 044013
  \bibitem{Higgs2} He M, Jinno R, Kamada K, Starobinsky A A and Yokoyama J J \emph{J. Cosmol. Astropart. Phys.} 2021 066
  \bibitem{Dragan_et_all} Dragan A, Dębski K, Charzyński S S, Turzyński K and Ekert A 2023 \emph{Class Quantum Grav.} \textbf{40} 025013
  \bibitem{Ehrlich} Ehrlich R 2022 \emph{Symmetry} \textbf{14} 1198
  \bibitem{Brukner} Brukner C Information theoretic foundations of quantum Theory (available at: \href{https://www.iqoqi-vienna.at/research/brukner-group/information-theoretic-foundations-of-quantum-theory}{www.iqoqi-vienna.at/research/brukner-group/information-theoretic-foundations-of-quantum-theory}) (Accessed on 11 November 2021).
  \bibitem{Stuckey} Stuckey W McDevitt T and Silberstein M 2022 \emph{Entropy} \textbf{24} 12
  \bibitem{Horodecki_QI} Horodecki R 2021 \emph{Acta Phys. Pol. A} \textbf{139} 197
  \bibitem{Heisenberg} Heisenberg W 1955 \emph{Niels Bohr and the Development of Physics} (Pergamon)
  \bibitem{Bethe} Salpeter E E Bethe H A 1951 \emph{Phys. Rev.} \textbf{84} 1232
  \bibitem{Path_integral} Takeuchi T 2019 Huygens' principle and Feynman's path integral (available at: \href{http://www1.phys.vt.edu/~takeuchi/Tools/CSAAPT-Spring2019-Huygens&Feynman.pdf}{www1.phys.vt.edu \\ /\~{}takeuchi/Tools/CSAAPT-Spring2019-Huygens\&{}Feynman.pdf})
  \bibitem{Hiesmayr} Bertlmann R A Grimus W Hiesmayr B C 2006 \emph{Phys. Rev. A} \textbf{73} 054101
  \bibitem{Einstein} Einstein A 1949 \emph{In Albert Einstein: Philosopher-Scientist}; P A Schilpp, (Open Court) pp 3–94
  \bibitem{Weinberg} Weinberg S 2017 The trouble with Quantum Mechanics (available at: \href{http://quantum.phys.unm.edu/466-17/QuantumMechanicsWeinberg.pdf}{http://quantum.phys.unm.edu/466-17/QuantumMechanicsWeinberg.pdf})
  \bibitem{Grudka} Grudka A and Wójcik A 2022 \emph{New J. Phys.} \textbf{24} 098001
  \bibitem{Santo} Del Santo F and Horvat S 2022 \emph{New J. Phys.} \textbf{24} 128001

\end{thebibliography}
\end{document}